 \documentclass{emulateapj}
 \usepackage{apjfonts, color}
\slugcomment{5 pages, 2 figures, accepted for publication in ApJL}

\shorttitle{Long-term Stability of X-ray Orbital Modulation in LS~5039}
\shortauthors{Kishishita et al.}

\begin{document}


\title{Long-term Stability of Non-thermal X-ray Modulation \\ 
in the Gamma-ray Binary LS~5039}


\author{Tetsuichi Kishishita\altaffilmark{1, 2}, Takaaki Tanaka\altaffilmark{3}, Yasunobu Uchiyama\altaffilmark{3, 4}, and Tadayuki Takahashi\altaffilmark{1, 2}}


\altaffiltext{1}{Institute of Space and Astronautical Science, Japan Aerospace Exploration Agency, Sagamihara, Kanagawa 229-8510, Japan}
\altaffiltext{2}{Department of Physics, The University of Tokyo, Bunkyo, Tokyo 113-0033, Japan}
\altaffiltext{3}{Kavli Institute for Particle Astrophysics and Cosmology, SLAC National Accelerator Laboratory
2575 Sand Hill Road M/S 29, Menlo Park, CA 94025}
\altaffiltext{4}{Panofsky Fellow}
%\altaffiltext{4}{Max-Planck-Institut f$\ddot{\rm u}$r Kernphysik, Saupfercheckweg 1, D-69117 Heidelberg}

\begin{abstract}
We report on long-term stability of X-ray modulation apparently synchronized with an  
orbital period of 3.9 days in the $\gamma$-ray binary LS~5039.
Recent observations with the {\it Suzaku} satellite in the year 2007, which covered continuously more than one orbital period, have provided us with detailed 
characterization of X-ray flux and spectral shape as a function of orbital phase. 
Motivated by the results from {\it Suzaku},
we have re-analyzed  the X-ray data obtained with {\it ASCA}, {\it XMM-Newton}, and {\it Chandra}  between 1999 and 2005, to investigate long-term behavior 
of LS~5039 in the X-ray band.
 We found that the modulation curves in 1999--2007 are surprisingly stable.
Even fine structures in the light curves such as spikes and dips are 
found to be quite similar from one orbit to another. 
The spectral characteristics observed in the past are consistent with those seen with {\it Suzaku} for some orbital phase segments. 
We suggest that magneto-hydrodynamical collisions between the relativistic outflow from a compact object and the stellar wind from the O star explain the clock-like non-thermal X-ray emission over eight years through remarkably stable production of high-energy particles near the binary system.
%The clock-like non-thermal X-ray emission over eight years 
%indicates that magneto-hydrodynamical collisions between the relativistic outflow from a compact object and the stellar wind from the O star result in remarkably stable production of high-energy particles near the binary system. 

%any scenario in which the X-ray and $\gamma$-ray emission are powered by 
%accretion onto a compact object. 
\end{abstract}





\section{INTRODUCTION}
 Gamma-ray binaries, a new sub-category of X-ray binaries that radiate 
very high energy  (VHE) $\gamma$-rays, have been attracting much attention 
since such sources were detected by modern air Cherenkov telescopes.
Up to now, TeV $\gamma$-rays are detected from four binary systems; 
LS~5039  \citep{aha05}, 
PSR B1259$-$63 \citep{aha05_psr}, Cygnus~X-1 \citep{albert06}, and 
LS I +61 303 \citep{albert07, acciari08}.
While a compact star powering the system is known to be a young radio pulsar in 
PSR B1259$-$63, a black hole in Cygnus~X-1 (still an evidence of TeV $\gamma$-ray detection), it remains controversial in the case of LS 5039 and LS I +61 303.
Also, HESS~J0632+057 is pointed out as a candidate to new gamma-ray binary \citep{hinton09}.

%and a magnetor in LS I +61 303 
%(though not yet firmly confirmed) \citep{bednarek09},

The LS~5039 system was designated as a high mass X-ray binary based on 
the {\it ROSAT} all-sky survey data by \cite{motch97}.
The stellar companion is a bright star with V=11.2 and its spectral type is O6.5V((f)).
The distance to the binary is estimated as $d=2.5\pm0.1$~kpc \citep{casares05}.
%Although a mass of a compact object is estimated to be $M\sim3.7~M_{\odot}$ {\bf by ref.}, its nature is not yet firmly established whether a black hole or a neutron star/pulsar. 
The compact object is orbiting around the stellar companion 
with an orbital period $P_{\rm orb}=3.90603\pm0.00017$ days and  with 
eccentricity $e=0.35 \pm 0.04$ \citep{casares05}.
Milliarcsecond-scale collimated outflows extending up to $\sim$1000~AU were detected by radio observations 
\citep{paredes00, paredes02, ribo08}.
%with a mildly relativistic velocity of $v\sim0.3~c$ 

Recent observations by H.E.S.S. revealed that LS~5039 is 
a source of Very High Energy (VHE) $\gamma$-rays and hence, is able to accelerate particles to 
multi-TeV energies \citep{aha05}. 
Moreover, H.E.S.S. detected an orbital modulation of the VHE emission with 
its orbital period \citep{aha06}.
The light curve as a function of orbital phase has a peak 
around  inferior conjunction 
(when the compact object appears between the optical star and the observer). 

Most recently, new X-ray observations with {\it Suzaku}
have enabled us to fully characterize the flux and spectral shape  
as a function of phase over one and a half orbital periods \citep{takahashi09}. 
The {\it Suzaku} observations 
showed that the X-ray spectrum at each orbital phase can be well fit by a power law
and that photon indices vary within $\Gamma = 1.45$--1.60 depending on orbital phase.  
Based on the {\it Suzaku} and H.E.S.S. data, it was concluded 
that the X-ray emission is likely to 
be synchrotron radiation from relativistic electrons 
which are also responsible for the VHE $\gamma$-ray emission 
via inverse Compton scattering \citep{takahashi09}.

Now that the X-ray characteristics of LS~5039 for more than one orbital period has been uncovered by {\it Suzaku}, 
it is crucial to compare the temporal and spectral behavior as a function of orbital phase between the {\it Suzaku} data 
and past X-ray data. 
Whether or not the orbital modulation is stable during the past decade should provide us with clues to understand 
the emission mechanism and the nature of the compact object in the LS~5039 system. 
In this Letter, 
we report on the discovery of 
the long-term stability of the X-ray orbital modulation, 
by 
combining the archival {\it ASCA}, {\it XMM-Newton}, and {\it Chandra} data with the recent {\it Suzaku} data.

\section{DATA REDUCTION}
LS~5039 has been observed several times in X-rays for limited period with {\it ASCA}, {\it XMM-Newton}, and {\it Chandra} \citep[see][]{vbosch07}. 
The log of those observations as well as the {\it Suzaku} observation is given
 in Table \ref{tab:log}.
Data reduction and analysis methods for the {\it Suzaku} data are described in \cite{takahashi09}.
Barycentric correction was performed for all the data in the following analysis.

\begin{deluxetable*}{ccllccccccl}
\tabletypesize{\small}
%\tabletypesize{\scriptsize}
%\rotate
\tablecaption{Results of Observations\label{tab:log}}
\tablewidth{0pt}
\tablehead{
\colhead{Date} & \colhead{} & \colhead{Mission} & \colhead{Observation ID} &
\colhead{MJD} & 
\colhead{Exposure (ks)} & \colhead{Phase} 
& \colhead{$N_{\rm H}$\tablenotemark{a,b}} & \colhead{$\Gamma$\tablenotemark{a}} 
& \colhead{Flux(1--10 keV)\tablenotemark{a,c}}
& \colhead{$\chi^2 ({\nu})$\tablenotemark{d}}
}
\startdata
1999--10--04&&  {\it ASCA} &47001000 & 51455.5& 24.4 & 0.30--0.49 & 6.4 $\pm$ 1.7 &1.55 $\pm$ 0.12 &9.59 $\pm$ 0.28& 5.7~(13)\\
&&&&&&0.30-0.40 &6.4 (fix) & 1.55 $\pm$ 0.10 & 8.73 $\pm$ 0.43 \\
&&&&&&0.40-0.49 &6.4 (fix) & 1.54 $\pm$ 0.07 & 10.42 $\pm$ 0.36\\
%2002--09--10&& {\it Chandra} & 2730& 52527.3& 10.6& 0.69--0.72 
%& \nodata & \nodata & \nodata \\
2003--03--08&&{\it XMM-Newton} &0151160201& 52706.3& 10.3 & 0.52--0.55& 7.2 $\pm$ 0.6 & 1.51 $\pm$ 0.07 & 9.20 $\pm$ 0.17 & 8.2~~(6)\\
2003--03--27&&{\it XMM-Newton} &0151160301 &52725.8 &10.3 & 0.53--0.56&  7.0 $\pm$ 0.5 & 1.45 $\pm$ 0.07 & 8.93 $\pm$ 0.17 & 5.9~~(6)\\
2004--07--09&&{\it Chandra}&4600&53195.1 & 11.0 & 0.67--0.71&7.9 $\pm$ 0.8 & 1.49 $\pm$ 0.06  & 12.33 $\pm$ 0.28 & 5.2~~(6) \\
2004--07--11&&{\it Chandra}&5341&53197.5 & 17.9 & 0.28--0.34& 7.9 $\pm$ 0.8 & 1.53 $\pm$ 0.05  & 7.33 $\pm$ 0.17 &  4.5~(10) \\
%2005--04--13&&{\it Chandra}&  6259&  53473.1 &5.0& 0.83--0.85
%& \nodata & \nodata & \nodata \\
2005--09--22&&{\it XMM-Newton}&0202950201 & 53635.8&15.1 & 0.46--0.51& 6.8 $\pm$ 0.4 & 1.43 $\pm$ 0.05 & 10.36 $\pm$ 0.15 & 8.9~(12)  \\	
2005--09--24&&{\it XMM-Newton}&0202950301 &53637.8 & 9.8 & 0.99--0.02& 7.0 $\pm$ 0.9 & 1.57 $\pm$ 0.10 & 5.99 $\pm$ 0.14  & 3.4~~(9) \\
2007--09--15&&{\it Suzaku}&402015010 & 54352.7 & 203.2 & 0.00--1.47 & 7.7 $\pm$ 0.2 & 1.51 $\pm$ 0.02 & 8.07 $\pm$ 0.06\\
\enddata
%% Text for table notes should follow after the \enddata but before
%% the \end{deluxetable}. Make sure there is at least one \tablenotemark
%% in the table for each \tablenotetext.
%\tablenotetext{a}{Sample footnote for table~\ref{tbl-1} that was generated
%with the deluxetable environment}
\tablenotetext{a}{Quoted errors are at the 90 \% confidence level.}
\tablenotetext{b}{$N_H$ units are $10^{21}~{\rm cm}^{-2}$.}
\tablenotetext{c}{Unabsorbed fluxes are given in units of $10^{-12}$~erg~${\rm cm}^{-2}$~${\rm s}^{-1}$. }
\tablenotetext{d}{$\chi^2$ is calculated from the modeled orbital light curve.}
\label{tab:ls5039_obs}
\end{deluxetable*}

%%%%%%%%%%%%%%%%%%
%% ASCA
%%%%%%%%%%%%%%%%%%
\subsection{{\it ASCA}}
The scientific payload of {\it ASCA} consisted of two co-aligned detector systems, the Gas Imaging Spectrometer (GIS; \citealt{ohashi96}) and the Solid-state Imaging Spectrometer (SIS; \citealt{burke91}).
The GIS and SIS detectors covered the energy range of 0.7--10~keV and 0.5--10~keV, respectively. Considering the low-background and large effective area at high energies, we used only the data from the GIS detectors. 
During the observation the GIS was operated in a normal PH mode, and standard screening procedures were applied for the analysis.
The source photons were accumulated from a circular region with a radius of 
$4\farcm 5$ while 
background region was chosen in the same field of view with the same radius and an offset of $18\arcmin$ from the source.
The averaged source and background count rates in the energy range of 1--10~keV are 0.12~cnt s$^{-1}$ and 0.02~cnt s$^{-1}$, respectively, for each GIS detector.
In order to improve statistics, the source photons of both GIS detectors (GIS2 and 3) were co-added.

%%%%%%%%%%%%%%%%%%
%% XMM-Newton
%%%%%%%%%%%%%%%%%%
\subsection{{\it XMM-Newton}}
For the {\it XMM-Newton} observations, we analyzed data from the European Photon Imaging Camera (EPIC), which consists of two MOS \citep{turner01} and one PN \citep{struder01} CCD arrays. 
All the observations were performed with the Medium filters; the observing modes were Prime Small Window (time resolution $\sim$6~ms) and Prime Partial W2, for 
the PN and MOS cameras, respectively.
Reduction and analysis of the data were performed following the standard procedure using the SAS 7.0.0 software package.
After discarding high-background time intervals, 
the source photons were accumulated from a circular region with a radius of 
$40\arcsec$.
Since the fluxes measured by the EPIC PN instrument show systematically $\sim10$--$15\%$ smaller values in comparison with the fluxes measured by the EPIC MOS, we used only MOS data and co-added source photons of both detectors (MOS1 and 2). The averaged source and background count rates in 1--10~keV are 0.45~cnt s$^{-1}$ and 0.02~cnt s$^{-1}$, respectively, for each MOS.

%%%%%%%%%%%%%%%%%%
%% Chandra
%%%%%%%%%%%%%%%%%%
\subsection{{\it Chandra}}
We analyzed data from Advanced CCD imaging Spectrometer (ACIS; \cite{garmire04}) for {\it Chandra} observations. 
Because of the superb angular resolution of the X-ray mirror 
($\sim0\farcs 5$ for on-axis observations), 
the {\it Chandra} data could be easily affected by pile-up, particularly when a source is observed on-axis. 
The pile-up effect causes underestimate of source flux and derivation of harder spectral index. 
In four data sets in the {\it Chandra} Data Archive, LS~5039 was observed on-axis for two of them (ObsID: 2730, 6259), 
and was observed off-axis for the other two (ObsID: 4600, 5341). 
In the previous analysis by \cite{vbosch07}, the spectra of the on-axis observations actually showed harder photon indices ($\Gamma\sim1.1$) 
compared with the {\it Suzaku} results. \cite{vbosch07} 
noted that their results are likely to be affected by the pile-up effects.  
We also analyzed the same data sets and concluded that the pile-up effects are significant and difficult to be accurately restored. 
We therefore made use of  the data from the two off-axis observations to avoid 
large systematic errors. 
In the data analysis, the {\it Chandra}
Interactive Analysis of Observations software package (CIAO 3.4) was used to extract the spectrum, following the standard procedure given in the {\it Chandra} 
analysis thread.
The source photons were accumulated from a circular region with a radius of 
$45\arcsec$ while 
the background region was chosen in the same field of view with the same radius and an offset of $2\farcm 5$ from the source.
The averaged source and background count rates in 1--10~keV are 0.40~cnt s$^{-1}$ and 0.02~cnt s$^{-1}$, respectively.

\begin{figure}
\epsscale{0.90}
%\plotone{./figures/draw_index_phasesynch_asca_090116.eps}
\plotone{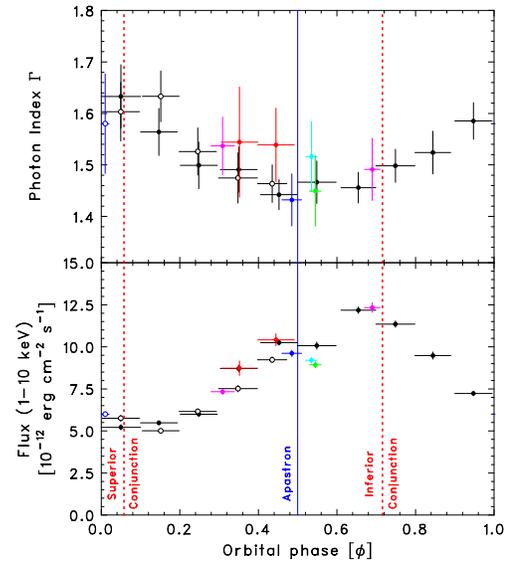}
\caption{
Orbital variations of photon index and unabsorbed flux in the energy range of 1--10~keV. Each color indicates the data of {\it XMM-Newton} (blue, cyan, and green), {\it ASCA} (red), and {\it Chandra} (magenta). Black filled and open circles represent the {\it Suzaku} observation.
Fitting parameters are shown in Table \ref{tab:ls5039_obs}.\label{fig1}}
\end{figure}

\begin{figure*}
\epsscale{0.878}
%\plotone{./figures/draw_allsatellite_flux_3figures.eps}
\plotone{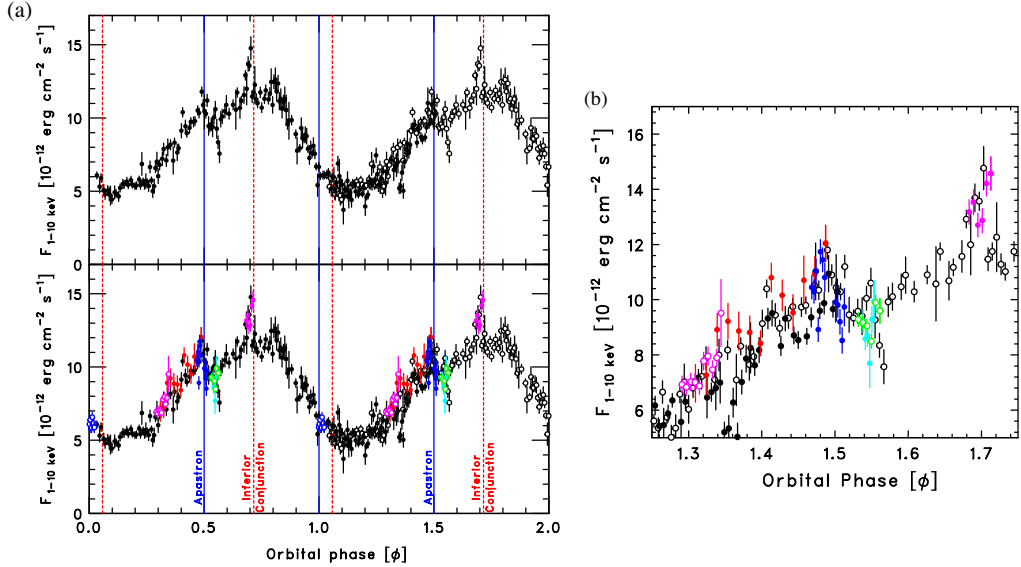}
\caption{(a) Orbital light curves in the energy range of 1--10~keV. (Top) {\it Suzaku} XIS data with a time bin of 2~ks. Overlaid in the range of $\phi=0.0$ to 2.0 is the same light curve but shifted by one orbital period (open circles). (Bottom) Comparison with the past observations. Each color corresponds to {\it XMM-Newton} (blue, cyan with each bin of 1~ks, and green with each bin of 2~ks), {\it ASCA} (red with each bin of 5~ks), and {\it Chandra} (magenta with each bin of 2~ks). Fluxes correspond to unabsorbed values. The blue solid lines show periastron and apastron phase and red dashed lines show {\it superior conjunction} and {\it inferior conjunction} of the compact object. (b) Closeup in $1.2\leq\phi<1.8$. \label{fig2}}
\end{figure*}

%%%%%%%%%%%%%%%%%%
%% Total
%%%%%%%%%%%%%%%%%%

% is shifted as is it would have been detected at the barycenter of the solar system instead of the satellite.

\section{ANALYSIS AND RESULTS}
\subsection{Spectral Analysis}
\cite{takahashi09} reported that the {\it Suzaku} spectra of LS~5039 are well fit with power laws 
and the photon indices vary within $\Gamma = 1.45$--1.60 depending on orbital phase
(see Figure \ref{fig1}).
We performed spectral fitting for the all 
archival data to compare  with the {\it Suzaku} results. 

All the spectra are well represented by absorbed power-law functions 
(using the $wabs$ model in XSPEC, \citealt{morrison83}). In Table \ref{tab:ls5039_obs}, we show the resulting fitting parameters, which agree with 
the literature (\citealt{vbosch07, martocchia05}).
The photon indices obtained for the {\it ASCA}, {\it XMM
-Newton}, and {\it Chandra} are within $\Gamma = 1.45$--1.55, and consistent with 
the {\it Suzaku} values within statistical errors. 
In addition, the $N_{\rm H}$ values obtained are generally consistent 
with the {\it Suzaku} best-fit values 
although the statistical errors are relatively large. 

In Figure \ref{fig1}, we plot 
the photon indices and  flux in the range 1--10~keV as 
a function of orbital phase. 
Here, the orbital phase is calculated with the period of 3.90603 days 
and $\phi=0.0$ with reference epoch $T_0$ (HJD $-$ 2400000.5 = 51942.59) taken from \cite{casares05}. 
Typical cross calibration uncertainty for photon index is 0.1, estimated from independent fits of Crab 
spectrum $\Gamma\simeq 2.1$ \citep{kirsch05,kokubun06}. 
We fixed the photoelectric absorption column density at $N_{\rm H}=0.77 \times 10^{22}
\ \rm cm^{-2}$ for phase-resolved {\it Suzaku} spectra, which is obtained 
from the phase-averaged {\it Suzaku} spectrum.
Also, phase-resolved {\it ASCA} spectra were fit using $N_{\rm H}$ frozen 
at the phase-averaged value. 
As seen in Figure \ref{fig1}, the photon indices obtained in the past observations follow the 
tendency seen in the {\it Suzaku} data, where the indices become smaller ($\Gamma \simeq 1.45$) around apastron and 
larger ($\Gamma \simeq 1.60$) around periastron.
TeV $\gamma$-ray emission presents similar trends in flux and photon indices \citep{aha06} as those of X-rays.
Remarkably, 
the X-ray flux shows almost identical phase dependency 
between the {\it Suzaku} and the other data sets as seen in the 
bottom panel of Figure \ref{fig1}. 
The similar tendency in fluxes and photon indices were also presented in \cite{vbosch05}, although the data were background contaminated and this prevented a proper photon index determination.
The results above indicate that the overall orbital modulation 
has been quite stable over the past eight years. 
Variability on shorter timescales is studied in the following subsection. 

\subsection{Temporal Analysis}
%%%%%%%%%%%%%%%%%%
%% Flux light curve
%%%%%%%%%%%%%%%%%%
The {\it Suzaku} light curves by \cite{takahashi09} revealed variability on
 short time scales of $\Delta \phi \lesssim 0.1$. 
To investigate the long-term behavior of the X-ray modulation, 
we compared the light curve by {\it Suzaku} and 
those obtained in the past observations.  
To directly compare the light curves obtained with the different detector systems, we need to convert detected counting rate 
to absolute energy flux for each bin of the light curves. 
For all the data other than the {\it Suzaku}, we assumed power-law spectra with parameters shown in Table~\ref{tab:log} 
and converted the counting rate of each time bin to power-law flux (unabsorbed flux). 
We adopted variable bin widths to have equalized errors for different data sets.
For the {\it Suzaku} data, since the spectral parameters, particularly photon indices, are significantly changed 
during the observation, we converted the observed counting rate assuming the power-law parameters obtained 
%for each phase interval of $\Delta \phi = 0.1$. \textcolor{blue}{In shorter time scales ($\Delta\phi\leq0.1$), short time feature appears in $0.62\leq\phi<0.73$ with significance from an average flux of 2.3 $\sigma$.
%%In each phase segment, significance of the short time features from an average flux varies from 1.1 ($0.83\leq\phi< $0.92) to 2.3 $\sigma$ ($0.63\leq\phi<0.73$).
%} 
The left panel of Figure \ref{fig2} shows the resulting  light curves in the energy range of 1--10~keV. 
Also, a magnified plot of the light curve is shown in the right panel of Figure \ref{fig2}. 
The phase-folded X-ray light curves from the \emph{ASCA}, \emph{XMM-Newton}, 
and \emph{Chandra} observations 
are in remarkable agreement with those from  {\it Suzaku}.
More surprisingly, short-time variability seen in the {\it Suzaku} 2007 data 
can be recognized in the previous 1999--2005  data. 
For instance, a small peak around $\phi = 0.70$ is evident both in the {\it Suzaku} and {\it Chandra} data 
although the two observations are performed almost  three years apart. When the {\it Suzaku} data points in $0.6\leq\phi<0.72$ excluding the peak structure are approximated to a linear function, the deviation of this peak structure from the continuum corresponds to 5.6 $\sigma$ significance.
The flux drop around $\phi=0.51$ is also consistent between 
the {\it Suzaku} and {\it XMM-Newton} data within statistical errors. The deviation of the drop structure from an extrapolated linear function determined from the {\it Suzaku} data points in $0.4\leq\phi<0.5$ is 7.4 $\sigma$ significance.
The {\it ASCA} data, which were obtained 8 years before the {\it Suzaku} observation, also reproduce 
the {\it Suzaku} light curve not only for the overall flux increase but also for the small peaks 
around $\phi = 0.40$ and $\phi = 0.48$. 
Since the X-ray spike around $\phi = 0.70$ is located around inferior conjunction (when the compact object appears between the stellar companion and the observer), it seems to be due to a rather geometrical than physical effect in the sense that it is connected to a direction to the observer. In this regard, the Doppler effect is a possible way of boosting or modulating the X-ray flux along the orbital phase, as pointed out in \cite{Bogovalov08}.
It should be noted that there are 
slight differences in the phase-folded light curves. 
Although a small spike at $\phi = 0.70$ in the {\it Suzaku} light curve 
agrees well with the  {\it Chandra} light curve as seen 
in the left panel of Figure \ref{fig2}, 
it is evident in the right panel that the {\it Chandra} spike 
lasts longer than the {\it Suzaku} one by $\Delta \phi \simeq  0.01$. 
Also, only the second orbit data of {\it Suzaku} shows a dip-like structure 
at $\phi = 0.35$, while there are no such structures in the other data 
including the first orbit data of {\it Suzaku}.

The largest systematic errors in flux of each time bin are due to uncertainties in the cross calibration 
between each detector system. 
Although cross calibration for a source with hard spectrum like LS~5039 has not been performed extensively, 
the discrepancy in the absolute flux is thought to be less than 20\% in the current calibration uncertainties (e.g. see http:\/\/www.iachec.org\/iachec\_2008\_meeting.html for the current cross calibration issues). 
This number is rather large if compared to the amplitude of the small structures 
seen in the light curves. 
However, the agreement of the similarity of shapes between {\it Suzaku} and other 
observatories would not simply be a chance coincidence in most cases since the agreement is seen 
for more than one observation by {\it Chandra} and {\it XMM-Newton}. 
In order to quantitatively estimate the similarity of shapes, we interpolated the {\it Suzaku} light curve and compared with each orbital segment's light curve.
The interpolated values are calculated with the {\it Suzaku} light curve of 5~ks time bins and a spline function, and $\chi^2$ values are derived from the difference between the interpolated and each light curves. Each $\chi^2$ value is pesented in Table \ref{tab:ls5039_obs}. The model light curve is acceptable in the confidence level of 10--90\% from $\chi^2$ test for each segment.
We note that the issue whether the source is really periodic or not at X-rays, despite the appearance, is still not solved statistically in periodicity analysis.
\section{DISCUSSION}
We have shown that the orbital modulation of LS~5039 is surprisingly stable 
over the past eight years. 
It is remarkable that not only the overall modulation but also the fine structures in the light curve agree well 
between the data from {\it Suzaku} and those from other X-ray observatories. 

%It is remarkable that not only the continuous sinusoidal component but also the wiggle-like structures agree with the {\it Suzaku} results.
%Since the X-ray light curves exhibit quite stable behavior over the past decades, it seems unlikely that the X-ray emission is an accretion origin.
%In the framework of the accretion scenario, the X-ray emission is related to the emission of the (Comptonized) hot accretion plasma formed around the compact object, or the miroquasar jet.
%In both cases, the instability of the accretion disk produces some variability in the light curve.
Considering the long-term stability, it is quite difficult to 
attribute the X-ray emission of LS~5039 
to the emission from hot plasma around an accretion disk, which is often observed from black hole binaries 
in low/hard state, since such emission is generally expected to show unstable 
and unpredictable time variability \citep{lewin06}. 
Although the jet formation and the production mechanism of non-thermal radiation in the jet are still unclear, it might be difficult to explain the emission from LS~5039 as synchrotron emission from a jet, since jet activities are thought to be connected to accretion activities and disk instabilities affect the jet formation and radiation \citep{fender04}.
Among currently proposed models for LS~5039, 
a scenario in which the radiation is related to an 
ultrarelativistic pulsar wind \citep{dubus06, sierpowski08} would be able to account for 
the stable clock-like X-ray and TeV $\gamma$-ray behavior. 
In this scenario, particles move relativistically in the pulsar wind, or are accelerated in shocks created by the interaction of this pulsar wind with the stellar wind. 
However, it should be noted that the H.E.S.S. results do not favor this scenario 
in its standard form, in which the emitter is expected to be located between the compact object and the stellar companion \citep{vbosch08}. 
In this case, the absorption is so severe that even cascading cannot help to explain disagreement between the measured and predicted fluxes around $\phi=0.0$.
In order to avoid the heavy absorption, $\gamma$-rays are assumed to be produced at the periphery of the binary system in \cite{takahashi09}, similarly as in \cite{Mitya_LS}.

%{\bf We need some statements about our modeling in Takahashi et al.\ (2009).
%Do we need to change the discussion in that paper? Probably not.
%Let's explain Takahashi et al.\ model. We would say the fine structures 
%are out of our scope in that paper.}

%The precise clock-like behavior rather favors a scenario in which the non-thermal radiation is related to an ultrarelativistic pulsar wind.
%In this scenario, particles are accelerated in shocks created by the interaction of a pulsar wind with the stellar wind. 
%\cite{takahashi09} reported that the X-ray modulation can be produced via an orbital change of adiabatic energy losses for accelerated electrons, which are dominant over radiation losses (synchrotron and inverse Compton scattering) in the binary system.
%This indicates that the adiabatic loss rate shows precisely periodic variability along the orbital period and/or other physical parameters (e.g. acceleration rate) are in the same condition. 

%A more complicated model has to be invoked to explain the VHE observations of PSR B1259-63 and LS I +61 303 possibly due to the presence of a Be star disk interacting with a compact object.
It is of importance to compare the results with those obtained for other gamma-ray binaries like PSR~B1259$-$63 and LS I +61 303. 
In contrast to the LS~5039 case, the X-ray emission from these two objects does not seem to show precise repeating features \citep{chernyakova06a, uchiyama09}. %\textcolor{blue}{Although the presence of the short time repeating features in the light curves is still unclear,} 
This might be due to the difference of the stellar object type or compact object in the systems. While LS~5039 host an O star, both PSR~B1259$-$63 and 
LS I +61 303 contains a Be star, which is characterized with a dense equatorial disk. 
Since a stellar disk plays an important role for X-ray emission in the pulsar scenario
\citep{tavani97}, it is possible that changes 
in disk structure avoid short time repeating features in X-ray modulation. \cite{romero07} actually pointed out that the stellar disk in LS I +61 303 would be likely disrupted around periastron passage by tidal forces from the compact object.
On the other hand, the emission from LS~5039 may be determined only by the  location of the compact object 
relative to the O star and thus exhibit precise repeating features.

%In conclusion, discovery of the periodic non-thermal X-ray emission in LS~5039 provides strong evidence of very stable acceleration processes occur in the binary system.

%%
\acknowledgments
We thank Felix Aharonian for fruitful suggestions and discussions. We also appreciate an anonymous referee for his/her helpful comments and suggestions to improve the manuscript. T. Kishishita is supported by research fellowships of the Japan Society for the Promotion of Science for Young Scientists.

\end{document}